\documentclass{aa}

\def\et{et al.}                        
\def\aap{A\&A}
\def\aaps{A\&AS}
\def\apj{ApJ}
\def\apjs{ApJS}
\def\aj{AJ}
\renewcommand{\deg}{$^\circ$}
\newcommand{\mtwc}[1]{ \multicolumn{2}{c}{ #1 } }

\begin{document}

   \thesaurus{
              03       
              (03.20.2 
              03.20.5  
              11.13.2  
              11.17.3  
              11.17.4  \object{3C\,345} 
              13.18.1  
               )} 
   \title{Total intensity and polarized emission of the parsec-scale 
          jet in \object{3C\,345}}

   \author{E.\ Ros
\and J.A.\ Zensus
\and A.P.\ Lobanov
          }

   \offprints{E. Ros, ros@mpifr-bonn.mpg.de}

   \institute{
Max-Planck-Institut f\"ur Radioastronomie,
Auf dem H\"ugel 69, D--53121 Bonn, Germany
             }

   \date{Received 23 September 1999 / Accepted 11 November 1999}

   \maketitle

\markboth{E. Ros, J.A.\ Zensus \& A.P.\ Lobanov}{Total intensity and
polarized emission of \object{3C\,345}}

\begin{abstract}
We have studied the 
parsec-scale structure of the quasar \object{3C\,345} 
at
$\lambda\lambda$\,1.3, 2, 3.6, and 6\,cm, using total intensity and linear
polarization images from 3 epochs of VLBA observations made in 
1995.84, 1996.41, and 1996.81.  The images show an unresolved 
``core" and a radio jet extending over distances up to 25\,milliarcseconds.
At 1.3 and 2\,cm, images of the linearly polarized emission reveal an 
alignment of the electric vector with the extremely curved inner 
section of the jet.
At 6\,cm, the jet shows strong
fractional polarization ($m$$\sim$15\%), with the electric vector 
being oriented
perpendicular to the jet.  This is consistent with the 
existence of a transition from the shock-dominated to the plasma 
interaction-dominated emission regimes in the jet, occurring at a distance
of $\sim$1.5\,milliarcseconds from the ``core".

   \end{abstract}


\section{Introduction\label{sec:introduction}}

The QSO \object{3C\,345} 
(V$\sim$16; $z$=0.5928, Marziani \et\ \cite{mar96})
presents one of the best 
examples of apparent superluminal motion in a core-dominated extragalactic
radio source, with components traveling in the parsec-scale jet along
curved trajectories at apparent speeds of
up to 10$c$ (Zensus \et\ \cite{zen95a}).  

\object{3C\,345} has been monitored using the 
Very Long Baseline Interferometry (VLBI) technique since
1979 (Unwin \et\ \cite{unw83}, Biretta \et\ \cite{bir86}, Brown \et\ 
\cite{bro94}, Wardle \et\ \cite{war94}, Rantakyr\"o \et\ \cite{ran95},
Zensus \et\ \cite{zen95a}, \cite{zen95b}, Lobanov \cite{lob96},
Taylor \cite{tay98}, Lobanov \& Zensus \cite{lob99}). 
On arcsecond scales the 
source contains a compact region at the base of a 4$^{\prime\prime}$ jet that
is embedded in a diffuse steep-spectrum halo (Kollgaard \et\ \cite{kol89}).  

From astrometric measurements, the
parsec-scale core has been shown to be stationary within
uncertainties of 20\,$\mu$as/yr (Bartel \et\ \cite{bar86}).
The parsec-scale jet consists of several prominent enhanced emission regions
(jet components) apparently ejected at different position angles (P.A.\
ranging from 240$^\circ$ to 290$^\circ$)
with respect to the jet core.  The components move along curved
trajectories that can be approximated by a simple helical geometry
(Steffen \et\ \cite{ste95}, 
Qian \et\ \cite{qia96}).
The curvature of the trajectories may be caused by 
some periodic process at the jet origin, e.g.\ orbital
motion in a binary black hole system (Lobanov \cite{lob96}) or 
Kelvin-Helmholtz instabilities.
At lower frequencies, the jet extends to the NW direction, turning northwards
at $\ge$20\,milliarcseconds (mas) distance from the core.

VLBI polarimetry of nonthermal radio sources
can provide stringent information about the physical conditions
in the parsec-scale jets.
Since the early works of Cotton \et\ (\cite{cot84}), 
substantial progress has been achieved in this area,
particularly benefiting from the enhanced observing capabilities
of the VLBA\footnote{Very Long Baseline Array, operated by the
National Radio Astronomy Observatory (NRAO).}.
The VLBA antennas have standardized feeds with low
instrumental polarization,
and use high-performance receivers.  
A new method of self-calibration of the 
polarimetric VLBI observations (Lepp\"anen \et\ \cite{lep95})
has made it feasible to measure the fractional polarization
with an accuracy of 0.15\%, by using
the target source itself as a calibrator.  
The technique is non-iterative and insensitive to the structure 
of the polarization calibrator.

In 1979--1993, \object{3C\,345} was
monitored using VLBI at several frequencies.
These observations have allowed the determination of
the trajectories and kinematic properties of the jet components 
(cf.\ Zensus \et\ \cite{zen95a}).  In this paper we extend this 
effort, and present three epochs of multi-frequency VLBA observations which
also include polarization information.
The observations were made in 1995.84, 1996.41, and 1996.81 at 
observing frequencies of
22, 15, 8.4, and 5\,GHz.
We describe the observations and data reduction in Sect.\ 
\ref{sec:observations}.  In Sect.\ \ref{sec:results},
we present the total intensity, polarization, and spectral index images of
the parsec-scale jet in \object{3C\,345}, and discuss the properties of the
jet emission.

Throughout this paper, we use a Hubble constant 
$H_0$=100$h$\,km\,s$^{-1}$\,Mpc$^{-1}$ and the deceleration parameter
$q_0$=0.5.  For \object{3C\,345}, this results in a linear scale of
3.79\,$h^{-1}$\,pc\,mas$^{-1}$.  A proper motion of 1\,mas\,yr$^{-1}$,
then,
translates into an apparent speed of 
$\beta_{\rm app}$=19.7\,$h^{-1}$$c$.
We use the positive definition of spectral index, $\alpha$ 
($S\propto \nu^{+\alpha}$).

\section{VLBA observations\label{sec:observations}}

We carried out multi-frequency observations of 
\object{3C\,345} at three epochs, 1995.84, 
1996.41, and 1996.81, using all 10 VLBA antennas.
The observations are summarized in Table~\ref{table:observations}.  
\object{3C\,345} was observed at each frequency with a 
5\,min scan every 20\,min for a total period of
almost 14\,h.  The bandwidth achieved in all cases was 16\,MHz.    
The data were registered in VLBA format, with an aggregate
data rate of 64\,Mbits/s, recording 4 channels with 1-bit sampling 
(mode 64-4-1, see Romney \cite{rom85}).
Some calibrator scans (the calibrator 
sources are listed in
Table~\ref{table:observations}) were inserted during the observations.

%
\begin{table*}[bthp]
\caption{\object{3C\,345} observation and map parameters}
\label{table:observations}
\begin{tabular}{@{}lclr@{$\times$}lrr@{$\pm$}lr@{$\pm$}lr@{$\times$(}l@{}}
        &    
              &          & \multicolumn{7}{c}{\object{3C\,345} image parameters} \\ 
        & Freq.\  
                  & \multicolumn{2}{c}{}  & & Beam   & \multicolumn{6}{c}{}    \\
Epoch   & $\nu$ 
                            & Calibrators 
                                       & \multicolumn{2}{c}{Beam size}  & P.A.      & \multicolumn{2}{c}{$S_{\rm tot}$$^{\rm a}$}  
                                                                                      & \multicolumn{2}{c}{$S_{\rm peak}$$^{\rm b}$} 
                                                                                        & \multicolumn{2}{c}{Contours in Fig.\ \ref{fig:all-maps}}\\
        & {\tiny [GHz] }
              &          & \multicolumn{2}{c}{{\tiny [mas]}} 
                                                       &                       & \multicolumn{2}{c}{{\tiny [Jy]}}
                                                                                       & \multicolumn{2}{c}{{\tiny [Jy/beam]}}
                                                                                                   & \multicolumn{2}{c}{{\tiny [(mJy/beam)$\times$(levels)]}}  \\ \hline
1995.84 & 22  
              & \object{3C\,279}  & 0.480 & 0.326 & --15$\rlap{.}^\circ$0 & 6.98&0.14 &  2.56&0.05 &  5.1 & --1,1,1.4,$\cdots$,256,362) \\
        & 15  
              & \object{3C\,279}  & 0.690 & 0.453 &  --0$\rlap{.}^\circ$7 & 7.98&0.16 &  3.66&0.07 &  7.2 & --1,1,1.4,$\cdots$,256,362) \\
        & 8.4 
              & \object{3C\,84}   & 1.254 & 0.731 &    2$\rlap{.}^\circ$9 & 9.32&0.18 &  4.46&0.09 &  6.7 & --1,1,1.4,$\cdots$,362,512) \\
        & 5   
              & \object{3C\,84}   & 2.443 & 1.348 & --21$\rlap{.}^\circ$0 & 7.51&0.15 &  4.41&0.09 &  4.4 & --1,1,1.4,$\cdots$,512,724.1) \\ \hline
1996.41 & 22  
              & \object{NRAO\,91} & 0.433 & 0.316 &  --3$\rlap{.}^\circ$1 & 4.51&0.09 &  1.80&0.04 &  3.7 & --1,1,1.4,$\cdots$,256,362) \\
        & 15  
              & \object{NRAO\,91} & 0.624 & 0.457 &  --7$\rlap{.}^\circ$3 & 6.70&0.13 &  2.67&0.05 &  5.3 & --1,1,1.4,$\cdots$,256,362) \\
        & 8.4 
              & \object{OQ\,208}  & 1.136 & 0.810 &  --7$\rlap{.}^\circ$9 & 7.49&0.15 &  3.55&0.07 &  5.3 & --1,1,1.4,$\cdots$,362,512) \\
        & 5   
              & \object{OQ\,208}  & 1.967 & 1.318 &  --9$\rlap{.}^\circ$5 & 7.21&0.14 &  3.91&0.08 &  3.9 & --1,1,1.4,$\cdots$,512,724.1) \\ \hline
1996.81 & 22  
              & --       & 0.451 & 0.322 & --15$\rlap{.}^\circ$5 & 4.61&0.09 &  2.09&0.04 &  4.2 & --1,1,1.4,$\cdots$,256,362) \\
        & 15  
              & \object{3C\,286}  & 0.614 & 0.474 &  --6$\rlap{.}^\circ$1 & 5.98&0.12 &  2.72&0.05  &  5.4 & --1,1,1.4,$\cdots$,256,362) \\    
        & 8.4 
              & \object{3C\,286}  & 1.132 & 0.887 & --16$\rlap{.}^\circ$8 & 7.52&0.15 &  3.59&0.07 &  5.4 & --1,1,1.4,$\cdots$,362,512) \\
        & 5   
              & \object{3C\,286}  & 1.861 & 1.532 & --15$\rlap{.}^\circ$3 & 7.26&0.15 &  4.29&0.09 &  4.3 & --1,1,1.4,$\cdots$,512,724.1) \\ \hline
\end{tabular}
\begin{list}{}{
\setlength{\leftmargin}{0pt}
\setlength{\rightmargin}{0pt}
}
\item[$^{\rm a}$] Total flux density recovered in the map model.  
The 
calibration errors are always smaller than a 2\%, which translates
into the tabulated values
for the uncertainties in $S_{\rm tot}$ and $S_{\rm peak}$.
\item[$^{\rm b}$] 
Peak of brightness in the map.
\end{list}
\end{table*}

%
\begin{figure*}[bthp]
\vspace{180mm}
\includegraphics{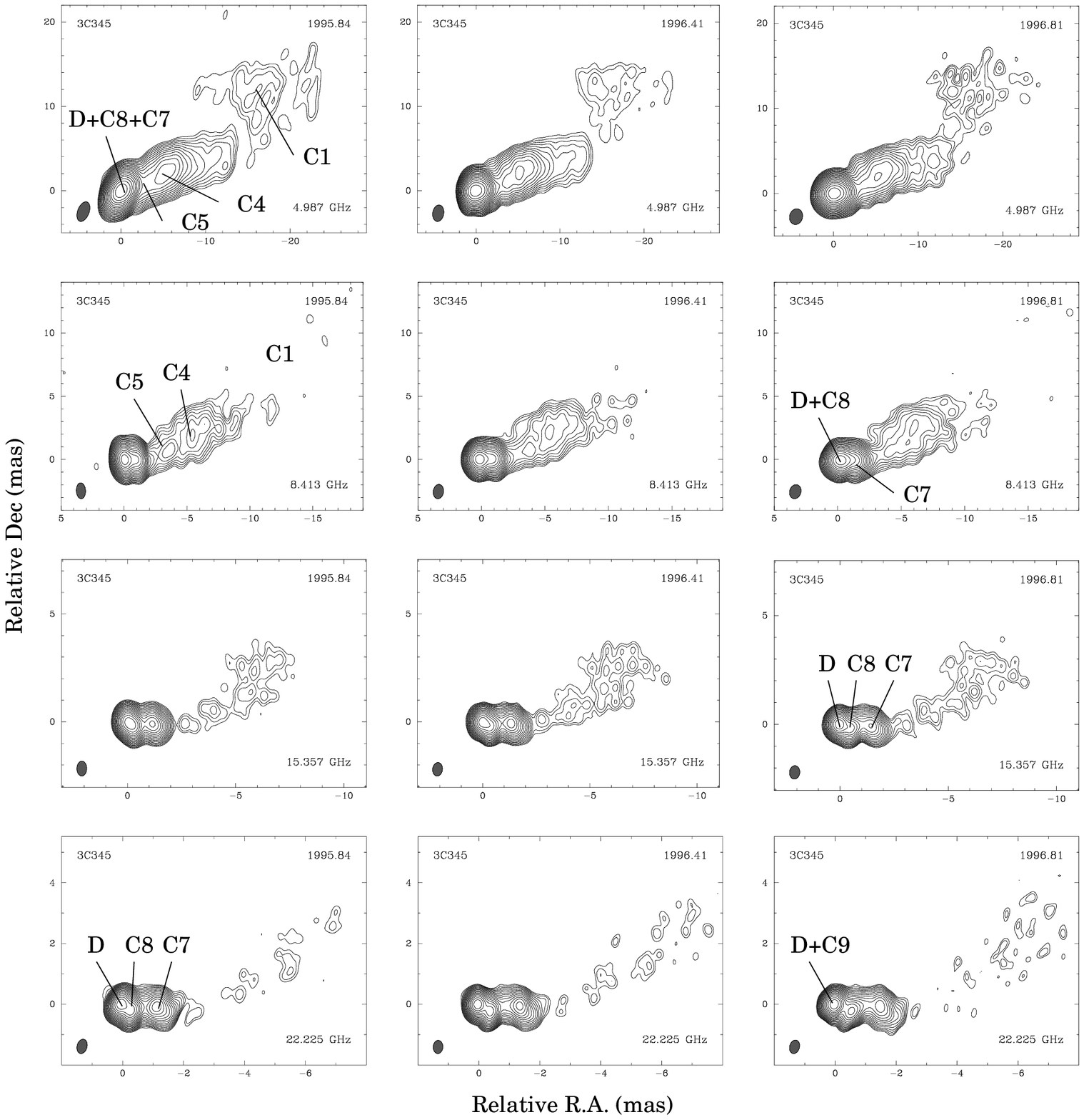}
\caption{VLBA total intensity flux
images of \object{3C\,345} for the three epochs and the four frequencies.  The 
map parameters are given on Table~1.  The synthesized interferometric beams
are represented at the bottom left of each image.  Notice that the
scales are not the same at different frequencies.}
\label{fig:all-maps}
\end{figure*}

The data were correlated at the NRAO VLBA Array
Operations Center in 
Socorro, NM, US.  A 2-s correlator pre-averaging time was used. 
Interferometer fringes were detected on all baselines for 
the target source and the calibrators.
The data were processed using {\sc aips}\footnote{Astronomical 
Image Processing System, developed and maintained by the NRAO.}.
The visibility amplitudes were calibrated using antenna gains 
and system temperatures measured at each antenna.  
The fringe fitting of the residual delays and fringe rates was performed 
independently for both parallel hands of the correlated
data, and the solutions obtained
were then referenced to a common reference antenna (Los Alamos for all cases).
After the fringe fitting, the data were averaged in frequency and
exported into {\sc difmap} (Shepherd \et\ \cite{she94}) for self-calibration
and imaging.  The data were further time-averaged, with the
averaging times selected individually at each frequency so as not to exceed
the respective coherence times.
We applied the {\sc clean} algorithm and self-calibration
in several cycles to produce the total intensity images.  

\subsection{Data analysis of the total intensity 
data\label{subsec:speproI}}

In order to quantify the source structure, we fitted models
consisting of elliptical Gaussian components to each of the self-calibrated 
data sets, using the tasks {\sc modelfit} and {\sc modfit} of the 
Caltech Package 
(Pearson \& Readhead \cite{pea88}).  For the purpose of studying the
spectral properties of the jet, we reproduced all images in a consistent
fashion, by selecting matching $uv$-ranges at all frequencies and
by restoring the final images with the same circular beam.  

\subsection{Imaging the polarized emission\label{subsec:imagP}}

Images of the linearly 
polarized intensity\footnote{$P=Q+iU=pe^{2 i \chi}=mIe^{2 i \chi}$, 
where $Q$ and $U$ are the Stokes linear polarization parameters,
$p=mI$ is the polarized intensity, $m$ is the fractional linear polarization,
and $\chi$ is the position angle of the electric vector (EVPA) in the sky.}
were obtained.
The instrumental polarization was determined using the feed solution algorithm
developed by Lepp\"anen \et\ (\cite{lep95}). 
The average accuracy of the antenna D-terms was better than 1\%. 
For some antennas, larger values of the D-terms were found at
higher frequencies: Brewster---4\% at 22\,GHz; OVRO---6--8\% 
at 22\,GHz and up to 4\% at 15\,GHz; Saint Croix---5\% in 
1996.81 at 22\,GHz.

We transferred the {\sc clean} component models from the total intensity ($I$) 
images back into {\sc aips} and then used the task {\sc imagr} to
produce the Stokes $U$ and $Q$ images of \object{3C\,345}.
The distribution of the polarized flux $p$ and the position angle
of the polarized emission was obtained from the combination of 
the $U$ and $Q$ images.
The registration of the $I$ and $P$ maps to within a small fraction of a
fringe spacing was assured by using the same set of antenna phases 
in the $U$ and $Q$ imaging.

The final step in the polarization calibration was to determine the absolute
offset between the right and left hand phases at the reference station.  
This can be done in one of three possible ways:
a) by observing a radio source with a known electric vector position 
angle $\chi$, 
b) by comparing the VLBI maps with VLA maps (that should account for
the total polarized intensity), and/or c) by comparing the
values of leakage of RCP into the left feed and LCP into the right feed
in the receivers (quantified as ``D-terms") 
of the antennas with values from neighbor epochs (relying in their 
time stability).  
We have applied the approaches b) \& c) in our data reduction.  For
the epochs and frequencies without such information, the right-
and left- hand offsets were introduced ad hoc, based on the consistency
with the calibration at other frequencies and epochs (for instance,
we were assuming that the outer, extended components cannot change their 
$\chi$ very rapidly from one epoch to another).  We summarize the
polarization calibration for all epochs and frequencies in 
Table~\ref{table:polar-calibration}.

%
\begin{table}[htbp]
\caption{Polarization calibration summary
\label{table:polar-calibration}
}
\begin{flushleft}
\begin{tabular}{@{}lc@{\,\,}c@{\,\,}c@{\,\,}c@{}}
        & 22\,GHz       & 15\,GHz       & 8.4\,GHz     & 5\,GHz      \\ \hline
1995.84 & {\footnotesize D-term$^{\rm a}$/VLA$^{\rm b}$} 
                        & {\footnotesize D-term.$^{\rm a}$/VLA$^{\rm b}$} 
                                        & {\footnotesize VLA$^{\rm b}$} 
                                                       & {\footnotesize VLA$^{\rm b}$} \\
1996.41 & {\footnotesize VLA$^{\rm b}$} 
                        & {\footnotesize Man.$^{\rm c}$}
                                        & {\footnotesize VLA$^{\rm b}$} 
                                                       & {\footnotesize Man.$^{\rm c}$} \\
1996.81 & {\footnotesize D-term$^{\rm d}$}
                        & {\footnotesize D-term$^{\rm d}$}
                                        & {\footnotesize Man.$^{\rm c}$}
                                                       & {\footnotesize Man.$^{\rm c}$}    \\ \hline
\end{tabular}
\begin{list}{}{
\setlength{\leftmargin}{0pt}
\setlength{\rightmargin}{0pt}
}
\item[$^{\rm a}$] Calibrated comparing with the D-terms at epoch 1995.83 (Alberdi, priv.\ comm.).
\item[$^{\rm b}$] Calibrated using VLA images.
\item[$^{\rm c}$] The right/left hand offsets were introduced manually, 
based on the consistency with the calibration at other frequencies and epochs.
\item[$^{\rm d}$] Calibrated comparing with the D-Terms at epoch 1996.85 (G\'omez, priv.\ comm.).
\end{list}
\end{flushleft}
\end{table}

At frequencies $\ge$5\,GHz, the ionospheric effects are negligible
and can be safely ignored.  We also do not apply Faraday
rotation corrections to the observed electric vector position angles,
relying on the results of Taylor (\cite{tay98}), who reports small 
rotation measurements ($RM=-130$\,rad\,m$^{-2}$ at the core and 
--70\,rad\,m$^{-2}$ at the parsec-scale jet, derived from a 
least-squares fit to the polarization angle measurements between 
8.1 and 15.2\,GHz) at frequencies higher than 5\,GHz.

\section{Results and discussion\label{sec:results}}

\subsection{Total intensity images and model 
fits\label{subsec:total-intensity-imaging}}

The total intensity images of \object{3C\,345} are shown in Fig.\ \ref{fig:all-maps}.
The parameters of the images are given in Table~\ref{table:observations}.  
The quality of the data and the performance of the VLBA antennas and
correlator were excellent.
The model fits for all images are presented in 
Table~\ref{table:modelfit-all}.

The errors in component position determined at each epoch are typically 
smaller than a tenth of a beamwidth, but to be conservative we have assumed
a magnitude of a fifth of beamwidth.  That means, approximatively, 0.08\,mas
for the 22\,GHz results, 0.11\,mas for 15\,GHz, 0.19\,mas for 8.4\,GHz, and
0.32\,mas for 5\,GHz.  The errors 
for the other parameters were determined combining the statistical 
standard errors provided by the task {\sc erfit} (revised and
improved version of {\sc errfit}) from the Caltech Package,
and some statistical and empirical considerations.

\subsubsection{The models at 22\,GHz}

%
\begin{table*}[ptbh]
\caption{Model fitting results} 
\label{table:modelfit-all}
\begin{flushleft}
\renewcommand{\baselinestretch}{0.75}
\begin{footnotesize}
\begin{tabular}{@{}ccr@{$\pm$}lr@{$\pm$}lr@{$\pm$}lr@{$\pm$}lr@{$\pm$}lr@{$\pm$}lr@{$\pm$}l@{}}\hline
     & 
     & \mtwc{\normalsize $S^{\rm a}$} & \mtwc{\normalsize $r^{\rm b}$} & \mtwc{\normalsize $\theta^{\rm c}$} & \mtwc{\normalsize $a^{\rm d}$} & \mtwc{\normalsize $b/a^{\rm e}$} & \mtwc{\normalsize $\phi^{\rm f}$} \\
     & 
     & \mtwc{\normalsize [Jy]} & \mtwc{\normalsize [mas]} & \mtwc{\normalsize [\deg]} & \mtwc{\normalsize [mas]} & \mtwc{} & \mtwc{\normalsize [\deg]} \\ \hline
&& \multicolumn{12}{c}{
---------------------
{\bf 22\,GHz}
---------------------
} \\ \hline
1995.84 &
Core (D) & 2.730 & 0.140  & \mtwc{---}      & \mtwc{---}    & 0.13 & 0.01 & \mtwc{1.0} & \mtwc{---}   \\
&
C8   & 2.252 & 0.110  & 0.40   & 0.08   & -122    & 11   & 0.25 & 0.01 & 0.66 & 0.01 & 47    & 1    \\
&
C7   & 1.214 & 0.060  & 1.20   & 0.08   &  -95    & 4    & 0.51 & 0.01 & 0.55 & 0.01 & 71    & 1      \\
&
Jet  & 0.400  & 0.020 & 6.20   & 0.14   &  -68    & 1    & 4.03   & 0.14   & 0.61 & 0.03 & 58    & 3      \\ \hline
1996.41 &
Core (D) & 2.168 & 0.110 & \mtwc{---}      & \mtwc{---}     & 0.16 & 0.01 & \mtwc{1.0} & \mtwc{---}      \\
&
C8   & 1.184 & 0.060 & 0.48   & 0.08   & -118    & 9    & 0.30 & 0.01 & 0.60 & 0.01 & -23 & 1      \\
&
C7   & 0.884 & 0.050 & 1.36   & 0.08   & -93     & 3    & 0.53 & 0.01 & 0.60 & 0.01 & 66  & 1      \\
&
Jet  & 0.334 & 0.017 & 5.21  & 0.11   & -73     & 2    & 6.8   & 0.2   & 0.33 & 0.02 & -64 & 1      \\ \hline
1996.81 &
Core (D) & 1.618 & 0.200 & \mtwc{---}      & \mtwc{---}     & 0.14  & 0.01  & \mtwc{1.00}  & \mtwc{---}    \\
&
C9   & 1.082 & 0.200 & 0.16   & 0.08   & -100    & 26    & 0.07  & 0.01  & 0.5  & 0.1  & 119   & 8      \\
&
C8   & 1.087 & 0.050 & 0.67   & 0.08   & -108    & 7    & 0.39  & 0.01  & 0.67  & 0.01  & 122   & 1      \\
&
C7   & 0.694  & 0.040 & 1.55   & 0.08   & -93     & 3    & 0.70   & 0.01   & 0.52  & 0.01  & 59  & 1    \\
&
Jet  & 0.230  & 0.012 & 5.9   & 0.1    & -72     & 4    & 5.0    & 0.2    & 0.5   & 0.2   & -51   & 2    \\ \hline 
&& \multicolumn{12}{c}{
---------------------
{\bf 15\,GHz}
---------------------
} \\ \hline
1995.84 &
Core (D) & 2.616 & 0.200 & \mtwc{---}   & \mtwc{---}    & 0.15  & 0.01  & \mtwc{1.0}  & \mtwc{---}      \\
&
C8 & 3.135 & 0.200 & 0.38   & 0.11   & -120   & 16  & 0.28  & 0.01  & 0.45  & 0.08  & -129    & 1    \\
&
C7 & 1.779 & 0.090 & 1.19   & 0.11   & -95    & 5   & 0.59  & 0.02  & 0.46  & 0.01  & 74    & 1    \\
&
Jet & 0.556 & 0.028 & 5.8    & 0.3    & -74    & 2   & 6.2   & 0.1   & 0.30  & 0.01  & -56   & 1      \\ \hline
1996.41 &
Core (D) & 2.501 & 0.130 & \mtwc{---} & \mtwc{---}          & 0.12  & 0.01 & \mtwc{1.0}  & \mtwc{---} \\
&
C8  & 2.227  & 0.110 & 0.46   & 0.11   & -115    & 13   & 0.32  & 0.01 & 0.85  & 0.01  & 7   & 1    \\
&
C7   & 1.544 & 0.080 & 1.36   & 0.11   & -93     & 5    & 0.60  & 0.01 & 0.58  & 0.01  & 66    & 1    \\
&
Jet   & 0.440 & 0.022 & 6.0    & 0.3    & -72     & 2    & 4.0   & 1.0   & 0.5  & 0.2  & 128   & 1      \\ \hline 
1996.81 &
Core (D+C9) & 2.844 & 0.140 & \mtwc{---} & \mtwc{---}        & 0.16  & 0.01 & \mtwc{1.0} & \mtwc{---}  \\
&
C8   & 1.762 & 0.090 & 0.57 & 0.11  & -108    & 11   & 0.38  & 0.01 & 0.79  & 0.01  & 97  & 1 \\
&
C7   & 1.017 & 0.050 & 1.48 & 0.11  & -93     & 4    & 0.70  & 0.01 & 0.50  & 0.01  & 59 & 1    \\
&
Jet  & 0.374 & 0.020 & 5.89  & 0.6   & -73     & 3    & 4.9   & 0.1   & 0.39  & 0.01  & -55  & 1    \\ \hline 
&& \multicolumn{12}{c}{
---------------------
{\bf 8.4\,GHz}
---------------------
} \\ \hline
1995.84 & 
Core (D) & 2.746 & 0.197 & \mtwc{---} & \mtwc{---} & 0.22 & 0.10 & \mtwc{1.0} & \mtwc{---} \\
& 
C8   & 2.337 & 0.185 & 0.33 & 0.19 & -119 & 30 & 0.21 & 0.02 & 0.72 & 0.08 & 106 & 22 \\
& 
C7   & 2.490 & 0.195 & 1.09 & 0.19 & -93 & 9 & 0.59 & 0.01 & 0.57 & 0.01 & 66 & 1 \\
& 
C5b & 0.082 & 0.020 & 2.35 & 0.3 & -92 & 10 & 1.0 & 0.2 & 0.5 & 0.2 & 100 & 9 \\
& 
C5a & 0.141 & 0.010 & 3.76 & 0.19 & -79 & 3 & 1.16 & 0.07 & 0.77 & 0.07 & 105 & 11 \\
& 
C4   & 0.552 & 0.028 & 6.29 & 0.3 & -69 & 7 & 3.0 & 0.2 & 0.7 & 0.1 & 132 & 2 \\
& 
C1   & 0.203 & 0.010 & 11.0 & 0.3 & -72 & 9 & 8.4 & 1.0 & 0.5 & 0.4 & 112 & 4 \\ \hline
1996.41 & 
Core (D) & 2.097 & 0.205 & \mtwc{---} & \mtwc{---} & 0.24 & 0.01 & \mtwc{1.0} & \mtwc{---} \\
& 
C8 & 2.304 & 0.115 & 0.43 & 0.19 & -115 & 24 & 0.32 & 0.01 & 0.59 & 0.06 & -27 & 2 \\
& 
C7 & 2.081 & 0.105 & 1.28 & 0.19 & -93 & 8 & 0.63 & 0.01 & 0.57 & 0.01 & 67 & 1\\
& 
C5 & 0.124 & 0.010 & 2.26 & 0.19 & -93 & 5 & 1.5 & 0.5 & 0.5 & 0.3 & 117 & 2 \\
& 
C4 & 0.713 & 0.036 & 6.07 & 1.0 & -71 & 8 & 4.36 & 0.03 & 0.50 & 0.01 & 126 & 1 \\
& 
C1 & 0.108 & 0.010 & 11.8 & 0.19 & -73 & 6 & 6.3 & 0.2 & 0.41 & 0.03 & 141 & 2 \\ \hline
1996.81 & 
Core (D) & 2.154 & 0.110 & \mtwc{---} & \mtwc{---} & 0.15 & 0.01 & \mtwc{1.0} & \mtwc{---} \\
&
C8 & 2.870 & 0.130 & 0.54 & 0.19 & -107 & 20 & 0.4 & 0.2 & 0.84 & 0.01 & 127 & 20 \\
&
C7 & 0.893 & 0.050 & 1.31 & 0.19 & -90 & 9 & 0.51 & 0.01 & 0.60 & 0.02 & 63 & 2 \\
&
C5b & 0.767 & 0.038 & 1.66 & 0.19 & -97 & 7 & 0.7 & 0.2 & 0.76 & 0.01 & 49 & 4 \\
&
C5a  & 0.171 & 0.015 & 3.75 & 0.19 & -84 & 8 & 2.45 & 0.04 & 0.5 & 0.1 & 106 & 1 \\
&
C4 & 0.556 & 0.030 & 6.60 & 0.19 & -69 & 6 & 3.4 & 0.5 & 0.66 & 0.01 & 119 & 1 \\
&
C1 & 0.131 & 0.010 & 12.17 & 1.0 & -72 & 7 & 7.0 & 2.0 & 0.6 & 0.2 & 141 & 5  \\ \hline 
&& \multicolumn{12}{c}{
---------------------
{\bf 5\,GHz}
---------------------
} \\ \hline
1995.84 & 
Core (D) & 1.892 & 0.345 & \mtwc{---} & \mtwc{---} & 0.15 & 0.10 & \mtwc{1.0} & \mtwc{---} \\
&
C8 & 1.695 & 0.328 & 0.37 & 0.32 & -112 & 41 & 0.49 & 0.01 & 0.4 & 0.2 & 87 & 3 \\
&
C7 & 2.394 & 0.130 & 1.08 & 0.32 & -90 & 17 & 0.63 & 0.01 & 0.4 & 0.2 & 64 & 1 \\
&
C5 & 0.189 & 0.010 & 3.20 & 0.32 & -82 & 6 & 2.2 & 0.1 & 0.4 & 0.2 & 97 & 4 \\
&
C4 & 0.800 & 0.040 & 6.30 & 0.32 & -69 & 3 & 3.58 & 0.03 & 0.54 & 0.01 & 120 & 1 \\
&
C1 & 0.305 & 0.020 & 17.6 & 0.5 & -59 & 2 & 17.2 & 0.6 & 0.26 & 0.02 & 152 & 2 \\ \hline
1996.41 &
Core (D) & 1.100 & 0.153 & \mtwc{---} & \mtwc{---} & 0.15 & 0.10 & \mtwc{1.0} & \mtwc{---} \\
&
C8 & 2.279 & 0.150 & 0.39 & 0.32 & -115 & 40 & 0.3 & 0.1 & 0.5 & 0.1 & -68 & 19 \\
&
C7 & 2.343 & 0.150 & 1.26 & 0.32 & -93 & 14 & 0.69 & 0.05 & 0.43 & 0.03 & 68 & 3 \\
&
C5 & 0.420 & 0.030 & 3.76 & 0.32 & -85 & 5 & 3.8 & 0.2 & 0.37 & 0.01 & 108 & 1 \\
&
C4 & 0.700 & 0.040 & 6.65 & 0.32 & -68 & 3 & 3.50 & 0.03 & 0.60 & 0.01 & 104 & 1 \\
&
C1 & 0.356 & 0.010 & 17.3 & 0.4 & -61 & 2 & 15.0 & 1.0  & 0.49 & 0.02 & 142 & 2 \\ \hline
1996.81 & 
Core (D) & 0.948 & 0.100 &  \mtwc{---} & \mtwc{---} & 0.15 & 0.03 & \mtwc{1.0} & \mtwc{---} \\
&
C8 & 2.929 & 0.150 & 0.49 & 0.32 & -112 & 33 & 0.49 & 0.01 & 0.66 & 0.01 & 93 & 1 \\
&
C7 & 2.025 & 0.100 & 1.42 & 0.32 & -94 & 13 & 0.84 & 0.01 & 0.47 & 0.01 & 64 & 1 \\
&
C5 & 0.347 & 0.020 & 4.03 & 0.32 & -82 & 5 & 3.44 & 0.02 & 0.4 & 0.2 & 104 & 1 \\
&
C4 & 0.677 & 0.040 & 6.79 & 0.32 & -68 & 3 & 3.57 & 0.02 & 0.49 & 0.01 & 104 & 1 \\
&
C1 & 0.368 & 0.020 & 16.7 & 0.5 & -62 & 2 & 17.7 & 0.3 & 0.36 & 0.01 & 141 & 1 \\ \hline 
\end{tabular}
\end{footnotesize}
\renewcommand{\baselinestretch}{1.0}
\begin{list}{}{
\setlength{\leftmargin}{0pt}
\setlength{\rightmargin}{0pt}
}
\item[$^{\rm a}$] $S$, flux density of the Gaussian component.
$^{\rm b}$ $r$, distance to the main component. 
$^{\rm c}$ $\theta$, position angle (defined north to west) of the component.
$^{\rm d}$ $a$, major axis at the FWHM of the elliptical Gaussian.
$^{\rm e}$ $b$, minor axis; $b/a$ ratio.
$^{\rm f}$ $\phi$, position angle of the major axis.
\end{list}
\end{flushleft}
\end{table*}

\begin{figure}[tbhp]
\vspace{68mm}
\includegraphics{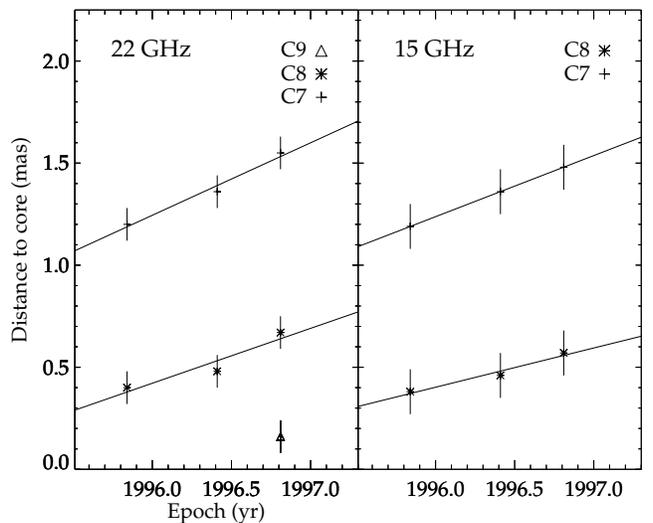}
\caption{Core separations for the jet components C7, C8, and C9 obtained from
the model fits to the data at 15 and 22\,GHz.
\label{fig:dist-22-15}}
\end{figure}

Table \ref{table:modelfit-all} lists the
Gaussian fits for the data at 22\,GHz.  We 
can model the radio source reliably with four components, that we 
associate with the core D, the components C8 and C7, and a more extended 
emission to the NW labeled as ``jet" (which contains the components
identified with C5 and C4 at lower frequencies).  Components C8 and 
C7 move away from
the core, at angular speeds of 0.27$\pm$0.12\,mas/yr (estimated from our three
observations, and corresponding
to an apparent speed of 5.3$\pm$2.3\,$h^{-1}\,c$, 
uncertainties
shown here and in the next subsections are statistical standard errors from
a weighted linear fit to the core distances)
and 0.35$\pm$0.12\,mas/yr 
(7.0$\pm$2.3\,$h^{-1}\,c$), respectively.  
In Fig.\ \ref{fig:dist-22-15}, we represent their proper motions 
by the changes of $r$ (distance to component D) with time.
The inner components of the jet of 3C\,345 have displayed complex trajectories
in previous epochs.  We cannot associate C8 with previous components,
since it is new, component C7 has been observed before and displays
acceleration from 1991 to our observing epochs.  With our time span
of a year we linearize the behavior of the C7 and C8 components to compute
their apparent velocities.  The comparison of the positions reported
here with later observing epochs at the same frequencies
should constrain the parameters of their kinematics.

%

A better fit is obtained at the epoch 1996.81, by introducing a new component 
between the core D and C8, labeled as C9, very close to D, 
at 0.16\,mas (P.A.\ $\sim -100^\circ$).  The model fitted fluxes
at D and C9 are strongly correlated. Different values for the
flux densities of D and C9 (preserving constant the sum of both) 
provide almost the same values for the agreement factor in the
fit.  To the first order of accuracy, the fluxes at the core and
C9 are close to $S_{\rm D}\sim1.6$\,Jy and $S_{\rm C9}\sim1.0$\,Jy. 

\begin{figure}[tbhp]
\vspace{125mm}
\includegraphics{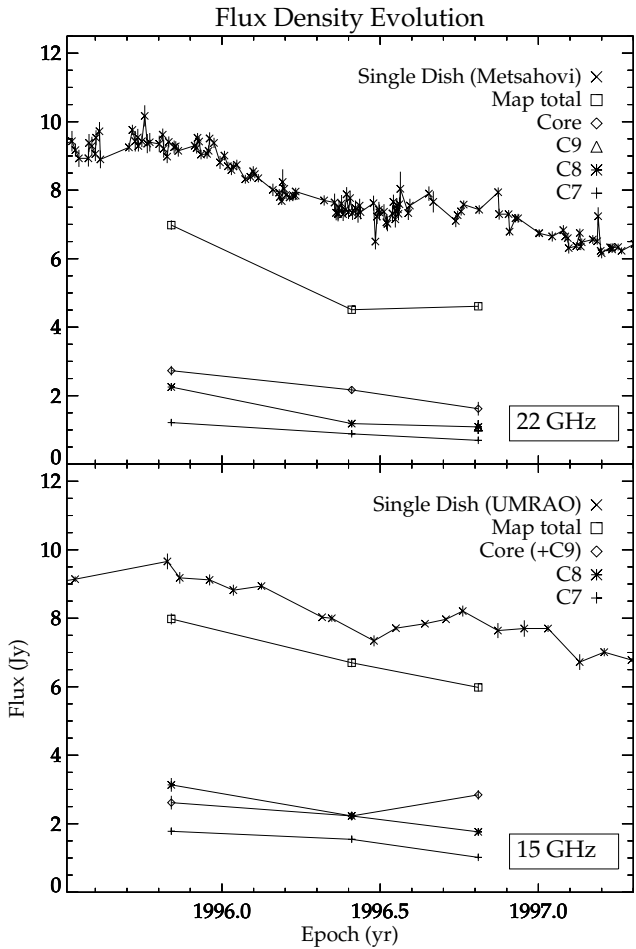}
\caption{Evolution of the flux densities at 22 
(Ter\"asranta \et\ \cite{ter98}) and 15\,GHz (Univ.\ of Michigan) 
for the single dish flux, the total VLBI flux (from Table 
\ref{table:observations}), and the fluxes of the model fit components (from 
Table \ref{table:modelfit-all}).
\label{fig:flux-mfit-22-15}}
\end{figure}

We show the flux density evolution of the different components in
Fig.\ \ref{fig:flux-mfit-22-15}, comparing also the values with the
total
flux density from the maps (Table \ref{table:observations}) and
the single dish measurements obtained in Mets\"ahovi 
(Ter\"asranta \et\ \cite{ter98},
and priv.\ comm.).  
Lobanov \& Zensus (\cite{lob99})
have applied a flare model to the observed variations of the
22\,GHz flux density of the core during our observations.  In their
description, the 
core is at the late stages of the decay after a flare in 1995.2.  
The new component C9 may therefore be related to the flare in 1995.2.
Similarly,
previous components have been associated to flares.  
The periods of emergence
of components are related to the periods of variability (see, e.g., Zhang 
\et\ \cite{zha98}).  This behavior is extremely clear for \object{3C\,273}
(T\"urler \et\ \cite{tue99}).
Following the work of Villata \& Raiteri (\cite{vil99}) 
on \object{Mkn\,501}, the most enhanced emission regions
in the jets (called ``components") would be 
the result of the boosting of helical features caused by the viewing angle.
The helicity would be 
generated by the rotation of a binary black hole 
system.  A similar model was discussed for
\object{QSO\,1928+738} by Hummel \et\ (\cite{hum92}).  The periodicity
in the helix would then be correlated with the light-curve periodicity in
the optical and radio, in this scenario.

%

\begin{figure}[tbhp]
\vspace{65mm}
\includegraphics{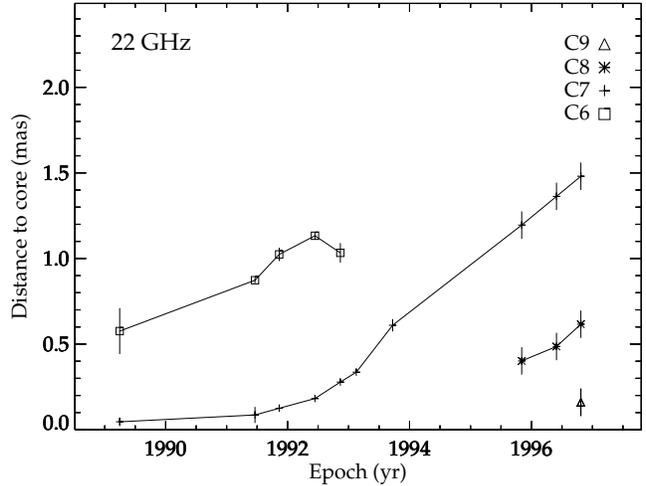}
\caption{Core separations of the jet components 
C6, C7, C8, and C9 at 22\,GHz obtained from
our model fits (last three epochs) and previous results 
(Lobanov \cite{lob96} and Lepp\"anen \et\ \cite{lep95}).  
The evolution and acceleration of
C7 is obvious.  The component C6 is not assigned in our model fitting,
and C8 and C9 are new with respect to the previous data sets.
\label{fig:dist-22}}
\end{figure}

The labeling of components follows Lobanov (\cite{lob96}).  
To identify our components with the trajectories reported before,
we have combined our component positions with previous ones.
We show in Fig.\ \ref{fig:dist-22} our data combined with
those mentioned above.  There we can track C7, see how C6 is not
detected now, and also the presence of C8 and C9.  The inspection
of the flux density values for C7 in previous data and our data
shows a continuity in the decay, from ~6\,Jy around 1992 to 
the values presented above.  The identification, thus, is
unambiguous.

%

\subsubsection{The models at 15\,GHz}

The model fitting of the source structure at 15\,GHz follows the 
same procedure, and provides results similar to the models at 22\,GHz.
Here, we also identify the core component D, two jet components C8 and
C7, and the extended component related to the faint emission of the
outer components of the jet.  We show the results in Table 
\ref{table:modelfit-all}.

The introduction of a new component
C9 at epoch 1996.81 between the core D and C8 does not improve the fit.
We plot the flux density changes in the lower panel of 
Fig.\ \ref{fig:flux-mfit-22-15},
comparing them with the total mapped flux density and single dish 
measurements from the University of Michigan Radio Astronomy Observatory
(hereafter UMRAO, see, e.g.,
Aller \& Aller \cite{all96}).

The proper motions of C8 and C7 measured at 15\,GHz 
(plotted in right panel of Fig.\ \ref{fig:dist-22-15})
are 0.19$\pm$0.16\,mas/yr 
(3.8$\pm$3.1\,$h^{-1}c$) and 0.30$\pm$0.16\,mas/yr (5.9$\pm$3.1\,$h^{-1}c$),
respectively.

\subsubsection{The models at 8.4\,GHz}

Model fitting of the source structure observed at 8.4\,GHz 
(and also at 5\,GHz) requires the inclusion of 
the inner components seen at the higher frequencies, 
and additionally of several more extended components at larger
distances from the core.
We present these resulting model fits in Table \ref{table:modelfit-all}.
Apart from the core D, C8 and C7, a strong component ($\sim$0.6\,Jy)
is found at $\sim$6.2\,mas (P.A.\ $\sim-70^\circ$) 
away of the core for the three epochs.  
It is most plausibly identified with the C4 component discussed by, e.g.,
Lobanov (\cite{lob96}).
The models are not robust for the region between C7 and C4, and
provide one or two components for different epochs.  For
consistency with other frequencies, we labeled it as C5 or
C5a and C5b (if double).

\begin{figure}[tbhp]
\vspace{65mm}
\includegraphics{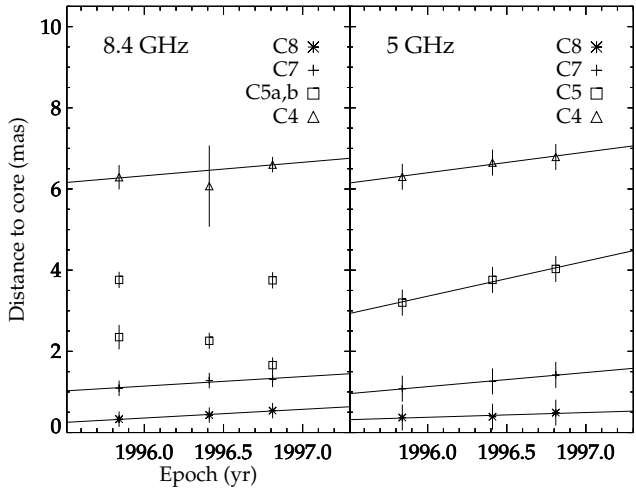}
\caption{Distance to core of components C8, C7, C5, and C4 at 8.4 and 5\,GHz from
our model fit. C5 splits into C5a and C5b in 
two epochs at 8.4\,GHz, where the
modeling does not provide robust fits to discern between 
one or two components.
\label{fig:dist-84-5}
}
\end{figure}

A new component is needed to fit also the emission at larger distances.
The extended emission to the NW at $\geq$10\,mas can be also reproduced
with an extended component (FWHM of $\sim$7\,mas), labeled as C1. 
The extended nature of these jet components makes proper
motion determinations very uncertain.  Tentative values would be,
for C8, 0.21$\pm$0.28\,mas/yr 
(4.2$\pm$5.4\,$h^{-1}c$), 
C7: 0.23$\pm$0.28\,mas/yr (4.6$\pm$5.4\,$h^{-1}c$),
and C4 with 0.33$\pm$0.37\,mas/yr (6.5$\pm$7.2\,$h^{-1}c$).  We display these proper
motions in Fig.\ \ref{fig:dist-84-5}.

%

\subsubsection{The models at 5\,GHz}

We modeled the visibility data at 5\,GHz with the same set of 
Gaussian components as in the models for 8.4\,GHz.  The results are 
presented in Table \ref{table:modelfit-all}.
The components in the core region D/C8/C7 are strongly correlated in 
flux density with each other, which limits the accuracy of the
flux densities. However, their positions can be derived accurately.  
The component C4 is evident at $\sim$6.5\,mas
(P.A. $\sim-70^\circ$).  The emission between the core region and C4
can be reproduced with a component at about 3.5\,mas out of the core.
Also, the most extended emission of the jet, 
is reproduced
with the component labeled C1, located at $\sim$15\,mas 
(P.A.\ $\sim-60^\circ$).   We show the evolution of the core separations
of C8, C7, C5 and C4
in Fig.\ \ref{fig:dist-84-5}.  Values for the
proper motions are 
C8: 0.12$\pm$0.46\,mas/yr (2.3$\pm$9.1\,$h^{-1}c$),
C7: 0.35$\pm$0.46\,mas/yr (6.8$\pm$9.1\,$h^{-1}c$), 
C5: 0.86$\pm$0.46\,mas/yr (17.0$\pm$9.1\,$h^{-1}c$), and
C4: 0.51$\pm$0.46\,mas/yr (10.0$\pm$9.1\,$h^{-1}c$).
(Note that the uncertainties presented are formal statistical standard 
deviations
from a weighted linear fit to the three points for every component.)


\subsection{Radio spectra\label{subsec:spectra}}

Lobanov (\cite{lob98b}) showed that it is possible to map 
the synchrotron
turnover frequency distribution using nearly simultaneous,
multi-frequency VLBI observations.  The turnover frequency
is defined as the location of maxima 
$(S_{\rm m},\nu_{\rm m})$, in the spectrum of the
synchrotron radiation from a relativistic jet
(spectral shape $S(\nu)\propto \nu^\alpha$).  The
spectrum is characterized by this maximum and the two
spectral indices, $\alpha_{\rm thick}$ ($\nu \ll \nu_{\rm m}$)
and $\alpha_{\rm thin}$ ($\nu \gg \nu_{\rm m}$).
Lobanov (\cite{lob98b}) showed the
first results from the imaging of the turnover 
frequency distribution of \object{3C\,345} from observations 
at 1995.48, at the same frequencies as those reported in
the present paper.

\begin{figure*}[tbhp]
\vspace{70mm}
\includegraphics{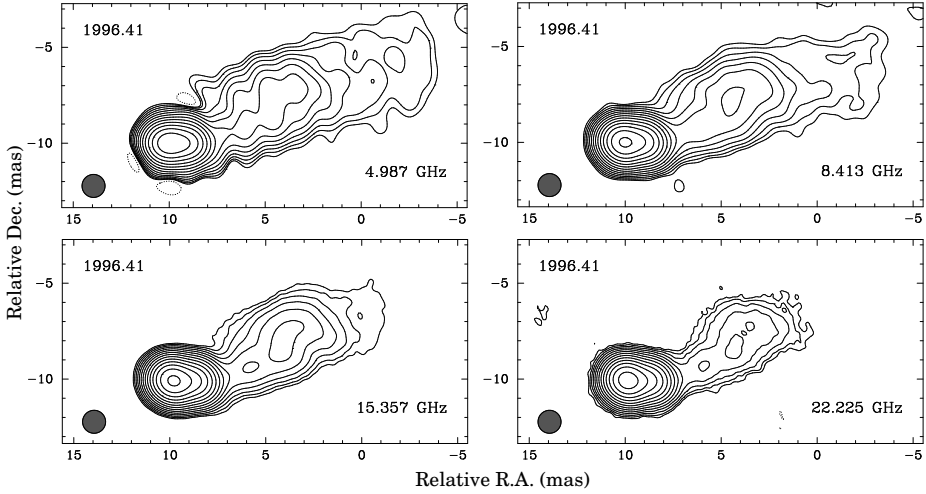}
\vspace{-60mm}
\hfill      \parbox[b]{5cm}{\caption[]{
Special renditions of the maps of \object{3C\,345} for epoch 1996.41 
with a circular restoring beam of 1.2\,mas diameter used to 
produce
the spectral index maps shown in figure \ref{fig:maps-spix1.2}.
The contours are 3\,mJy$\times$(-1,1,1.73,3,...,).  The peaks of brightness
are of 3.69, 4.24, 4.30, and 3.05\,Jy/beam respectively at 5, 8.4, 15, 
and 22\,GHz.  The shift of (10,-10\,mas) with respect to the maps
of Fig.\ \ref{fig:all-maps} was made to reduce map sizes
by computing reasons (the shift is unimportant as relative right
ascension and declination are shown).
Epochs 1995.84 and 1996.81 show similar features.
\label{fig:maps-1.2}
} 
}
\end{figure*}

\begin{figure*}[tbhp]
\vspace{171pt}
\includegraphics{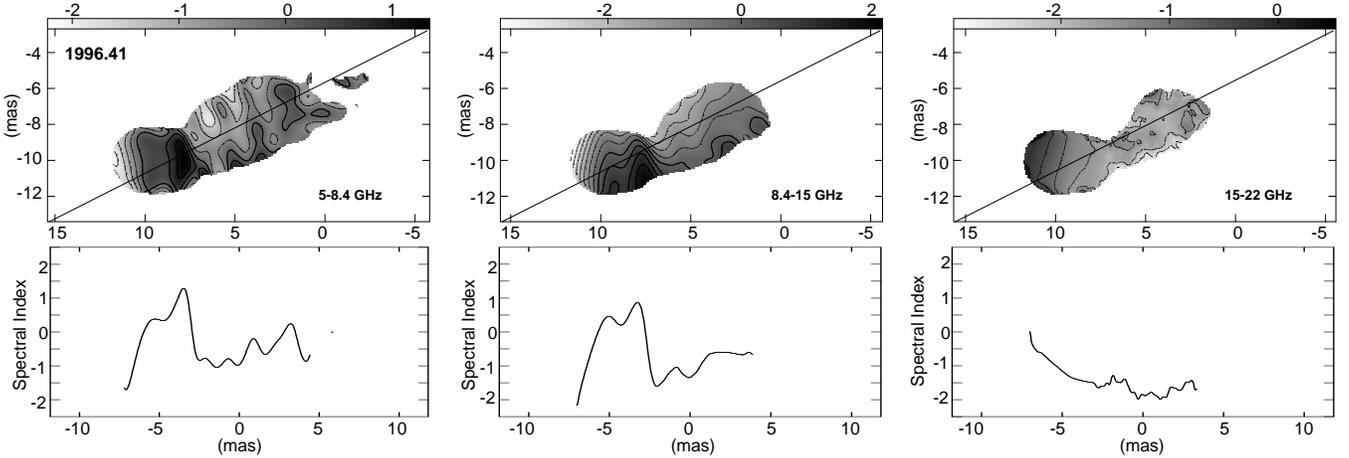}
\caption{Spectral index maps (top in every subimage) and profiles (bottom) 
among the maps convolved with a 1.2\,mas circular restoring beam 
(Fig. \ref{fig:maps-1.2}) from our multifrequency data set at 1996.41.
The profile line follows the diagonal line over the maps, with a P.A.\ 
of 117$^\circ$.
\label{fig:maps-spix1.2}
}
\end{figure*}

We produced special renditions of our maps at all
frequencies with a circular convolving
beam of 1.2\,mas in diameter (see Fig.\ \ref{fig:maps-1.2}),
in order to produce spectral index maps of \object{3C\,345}
(Fig.\ \ref{fig:maps-spix1.2}).  We display only the images
for epoch 1996.41.  In Fig.\ \ref{fig:maps-spix1.2}
we also show cuts through the spectral index maps along a line with a P.A.\ of
117$^\circ$.  Epochs 1995.84 and 1996.81, which are not represented
for brevity, show very similar features.  

%

%

The registration of the images was done by considering the relative 
separations of the model fit components from the core (assuming that the
components are optically thin), 
the theoretical predictions for the core registration
(see Lobanov \cite{lob98a}), and the alignment of the jet components 
directly in the images.  

In the spectral index maps, two different regions in the parsec-scale
structure of \object{3C\,345} are evident.  
The D/C8/C7 region is optically thick.  
The region of the more extended jet from
about $\geq 1.5$\,mas out of the core to the NW (P.A.\ $\sim -70^\circ$)
is optically thin.  The transition between both areas is located to
the west of the component C7.

The core region displays spectral indices that allow one to deduce that the 
turnover frequency at this region is above 15\,GHz, peaking between the
core and C8, and decreasing to the west.  The component
trajectories obtained exhibit apparent proper motions of
0.2--0.3\,mas/yr.  The peaks of the spectral images are consistent
with a shock evolution.  
This behavior and the shock parameters have been studied in detail 
by Lobanov \& Zensus (\cite{lob99}) for previous observing epochs.

In the comparison of data sets of 5/8.4 and 15/22\,GHz we find a 
gradient of the spectral index, increasing to the south: the southern 
edge of the jet is optically thicker than the northern
edge.  This has been observed at all three epochs, and we 
reject the possibility of a registration error after a careful
test of different shift values.  In the maps at 22\,GHz,
we observe a steeper fall in the flux 
density from the components to the south than to the north.  
There is some
extended emission in this northern region.  These gradients can
be interpreted as an effect arising from changes 
in the speeds across the jet flow.
The curved trajectories of the components can also be related to
this effect.

At larger distances along the
parsec-scale jet ($>$1.5mas),
the spectrum
shows an optically thin region, with the spectral index decreasing,
becoming as steep as
$\alpha_{\rm 22\,GHz}^{\rm 15\,GHz}\sim-2$.  The peaks in the
maps that include 5\,GHz information can 
also be affected by beam overresolution.
A similar gradient to the south is also evident in this
region in most of the cases.  
There seems to be an intrinsic asymmetry
in the jet, as we will also see in the polarization analysis, in 
Sect.\ \ref{subsec:linear-polarization}.
For a helical jet geometry (e.g., Villata \& Raiteri \cite{vil99}) 
this intrinsic asymmetry can be explained by the
beaming of emission towards the observer for one side of the helix ---the
southern in our case, and
fainter emission in the other side.

In summary, the spectral behavior of the jet shows a scenario with a clear
transition between two zones with different physical properties: the
core region and the outer parsec-scale jet region.  The components in
the latter show a slower flux decrease than in the inner region (comparing
the values in the model fit), and also higher speeds of the component
proper motions.

\subsection{Polarimetric imaging\label{subsec:linear-polarization}}

\begin{figure}[htbp]
\vspace{73mm}
\includegraphics{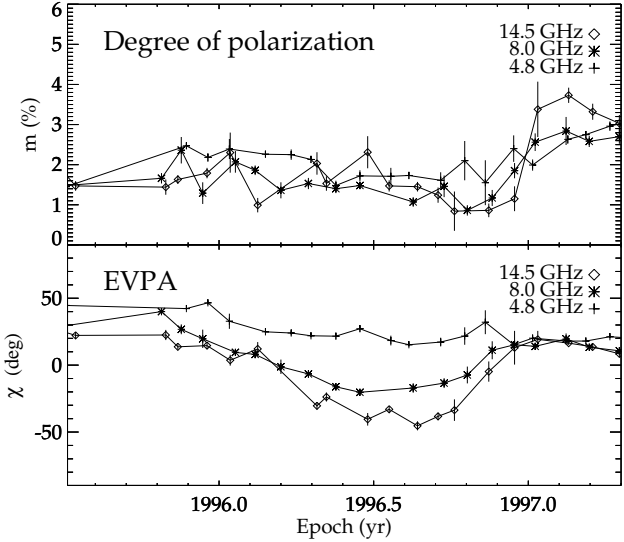}
\caption{Single dish $m$ and $\chi$ of \object{3C\,345} (UMRAO).
\label{fig:chi-umrao}
}
\end{figure}

Before discussing the VLBI results, we want to point out
some facts observed in the single-dish monitoring results of the 
polarized emission of \object{3C\,345}.
In Fig.\ \ref{fig:chi-umrao}, we plot the 
degree of polarization $m$ and the EVPA $\chi$ from the UMRAO data at epochs 
close to our observations.

%

The polarization angle of the total emission
$\chi$ shows a value of $\sim 40^\circ$ close
to 1995.84 and is constant around $\sim 20^\circ$ after early 1996.  
At higher frequencies, $\chi$ changes, showing a decrease and a further
increase in its value.  $\chi^{\rm UMRAO}_{\rm 15\,GHz}$ changes from 
values of $\sim 10^\circ$ at 1995.8 to $\sim -40^\circ$ in 1996.6,
turning back to $\sim 20^\circ$ on early 1997.  The degree of polarization
remains between 1--2\% during our three observing epochs.  Interestingly,
it increases up to 4\% after our last VLBI observations in 1997.1, when
the values $\chi$ at the three frequencies coincide again.  This is the epoch
where the C9 component appears in our model fitting, and also when the 
flux density begins to increase at the higher frequencies.

We can assume that the single dish polarization measurements of \object{3C\,345}
are almost equivalent to integrating all of the parsec-scale jet polarized
components.  The VLA polarized images mentioned above are consistent 
also with the single-dish measurements.  At the epoch of 1995.8, 
$\chi^{\rm VLA}_{\rm 15\,GHz} \sim 5^\circ$, and 
$\chi^{\rm VLA}_{\rm 22\,GHz} \sim -10^\circ$.

\begin{figure*}[htb]
\vspace{78mm}
\includegraphics{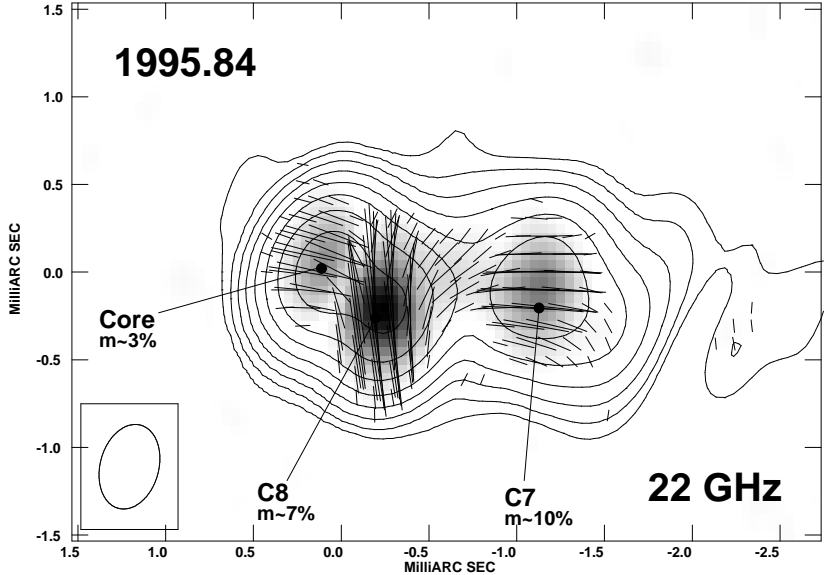}
\vspace{-50mm}
\hfill      \parbox[b]{5cm}{\caption[]{
\label{fig:pol22}
VLBA $I$, $p$ and $\chi$ images of \object{3C\,345} at 22\,GHz,
epoch 1995.84.  The total intensity $I$ is represented with contours 
(value of 6\,mJy/beam$\times\, -1,1,2.24,5,11.18,25,\dots)$, 
superimposed over a grey 
scale polarized intensity map (peak of brightness of 112.4\,mJy)
and the superimposed electric vectors ($\chi$, length proportional
to $p$, 1\,mas in
the map is equivalent to 100\,mJy/beam).
}
}
\end{figure*}

In Fig.\ \ref{fig:pol22}, we show the composition of the total intensity 
$I$ (contours), the polarized intensity $p$ (grey scale) and the electric 
vector orientation angle $\chi$ (segments, length
proportional to $p$) for the observations in 1995.84 at 22\,GHz.
It is obvious that the electric vector is 
aligned with the extremely curved jet direction in the inner 3\,mas.

%

\begin{figure}[tbhp]
\vspace{155mm}
\includegraphics{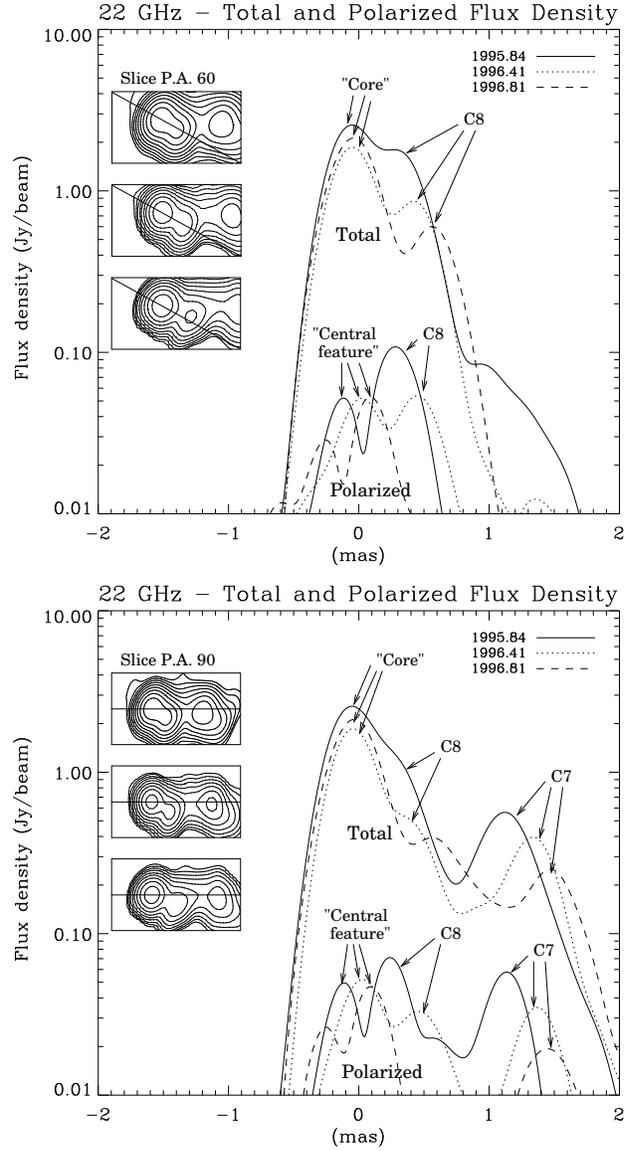}
\caption{Logarithmic slice profile of the total and polarized flux density maps
of \object{3C\,345} at 22\,GHz through the core and the C8 (top) and C7 (bottom)
components, crossing 
the center of the maps with a P.A.\ 60\deg and 90\deg, respectively (shown
at the top left of each figure).
The motion of the components in total intensity can be appreciated,
and also the changes in the polarized flux density features.
\label{fig:k-icln-poli}
}
\end{figure}

Fig.\ \ref{fig:k-icln-poli} shows the 
total and polarized flux density for the
three epochs along two cuts through the core, which are oriented at 
P.A.\ of 60$^\circ$ and 90$^\circ$, which cross, respectively, the 
components C8 and C7.  
The evolution of the total flux for the core
D and components C8 and C7 is evident in the figure.
The brightness peak of the polarized emission
moves westwards, and a new component appears in
1996.81.  The evolution of C8
is the most dramatic: in 1995.84, it dominates the $p$-map, then
it becomes much weaker 
in 1996.41 and vanishes in 1996.81.  The polarization
angle of C8 rotates from the 
1st to the 2nd epoch.  The component C7 shows a westward 
motion, similar in total and polarized emission.  The value of $\chi$
for this component is nearly constant throughout the period of
observations.
We can compare those results with the 22\,GHz map from \object{3C\,345}
at epoch 1994.45 presented in Lepp\"anen \et\ (\cite{lep95}), 
1.39\,yr
before our first epoch.  Component C8 had emerged and was clearly
visible in the polarization map, but not in the total intensity image.
Its EVPA began to be oblique with respect to the values of D and C7,
which were parallel to the jet, as it is observed also in our 
data.  Our maps at 22\,GHz represent a later stage in the evolution
of the C8 component, which is rotating through its evolution.

%

G\'omez \et\ (\cite{gom94}) have modeled features
similar to C8 from simulations of the emission in a bent shocked
relativistic jet, reporting
anticorrelation between the polarized and the total intensity emission,
which is observed at early stages of evolution of a relativistic shock.
We should point out that a possible superposition between components with 
different electric vector orientations may lead to cancellation
of the polarized flux and can produce apparent 
gaps between polarized emission components.  The brighter polarized 
emission in C8 might be explained by a shock wave in a curved jet.
More generally, the features in Fig.\ \ref{fig:pol22} can be explained in
terms of the 
shock model (Wardle \et\ \cite{war94}) 
in the framework of a helical geometry for the motion of the components 
(Steffen \et\ \cite{ste95}, Villata \& Raiteri \cite{vil99}).
Lister \et\ (\cite{lis98}) report similar vector alignment with the 
jet at 43\,GHz, for
the inner jet regions in several blazars.

\begin{figure}[htb]
\vspace{120mm}
\includegraphics{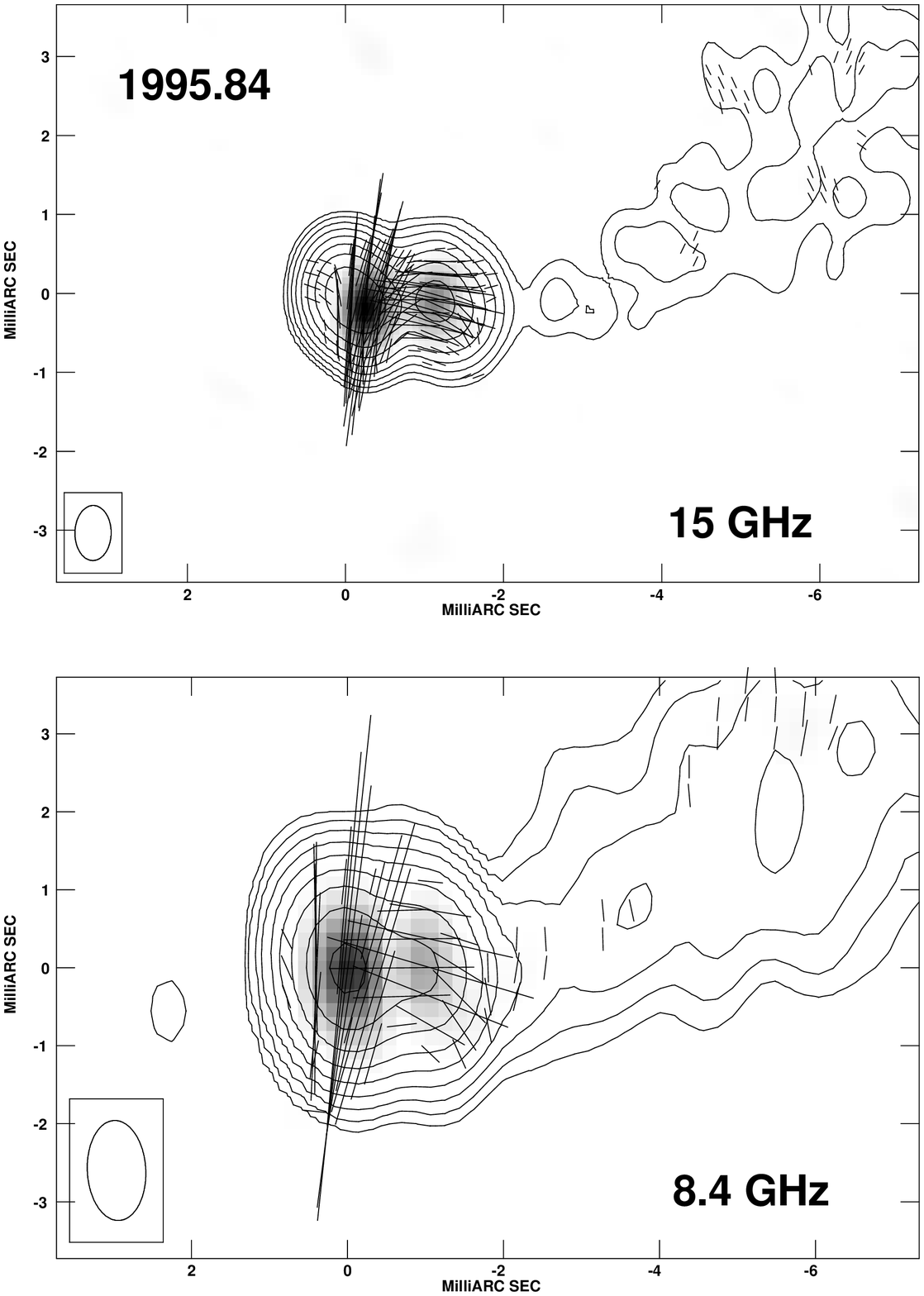}
\caption{VLBA $I$, $p$ and $\chi$ images of \object{3C\,345} at 15 and 8.4\,GHz,
epoch 1995.84.  The total intensity $I$ is represented with contours 
(value of 6\,mJy/beam$\times\,-1,1,2.24,5,11.18,25,\dots$), 
over a grey scale polarized intensity map (peaks of brightness of 
169.3 and 141.4\,mJy/beam, respectively, grey scale 
normalized to the common maximum polarized flux density --at 15\,GHz--),
and the electric vectors ($\chi$, length proportional to 
$p$, 1\,mas in
the map is equivalent to 50\,mJy/beam).
\label{fig:pol15-8.4}
}
\end{figure}

In Fig.\ \ref{fig:pol15-8.4}, we display the images of polarized emission
in \object{3C\,345}
at 15 and 8.4\,GHz, following the same scheme as in Fig.\ \ref{fig:pol22}.
The same features seen at 22\,GHz are present here.  At 8.4\,GHz, weak
polarized emission is also detected in the outer jet, at the northern
edge, with $\chi \sim 0^\circ$.  This change of the electric vector,
roughly parallel to the jet direction at D and C7 contrasts with its
value roughly perpendicular to it at the edge of the outer jet.  
The spectral analyses have shown that the jet at distances $>$1.5\,mas
of the core is an optically thin region,
where also a north-south gradient in $\alpha$ was present.

%

\begin{figure*}[hbt]
\vspace{75mm}
\includegraphics{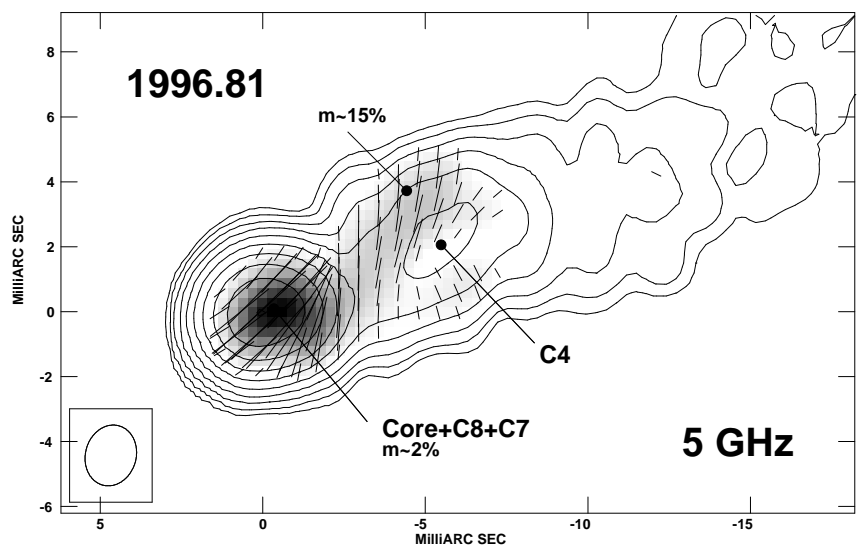}
\vspace{-55mm}
\hfill      \parbox[b]{5cm}{\caption[]{
Polarized intensity electric vectors ($\chi$, length proportional to $p$, 
1\,mas in the map is equivalent to 10\,mJy/beam)
overlaid on total
intensity ($I$) contours (3\,mJy/beam$\times\,-1,1,2.24,5,11.18,25,...$)
and grey scale polarized intensity ($p$, grey scale up to the peak of
brightness, 40.5\,mJy/beam) images for \object{3C\,345} at 5\,GHz, epoch 1996.81.  
It is obvious that
the electric vector is almost perpendicular to the jet at core separations
from 3 to 7\,mas.
\label{fig:pol5}}
}
\end{figure*}

At 5\,GHz, the distribution of the polarized emission differs 
significantly from that seen at higher frequencies.
In Fig.\ \ref{fig:pol5},
we present the image obtained for the 3rd epoch of our monitoring.
The two earlier epochs display very similar
features in the jet. 
The core is less polarized ($m\sim1\%$) than at higher frequencies,
with its electric vector oblique to the jet direction.  Both
facts may be caused by the blending of two or more emission regions
with different polarization angles.
The degree of polarization at 5\,GHz in the jet reaches 
$\sim$15\%.  
The predominance of the
polarized emission at the jet edges (especially at the northern 
one) is more obvious here than at 8.4\,GHz.

%

These findings seem to be consistent with 
the conclusions of
Brown \et\ (\cite{bro94}), who reported, at 5\,GHz, a weakly polarized 
core in \object{3C\,345} and a fractional polarization reaching 15\% in the jet.  
The electric vector position angles were 
found to be variable in the core from one epoch to the other but 
perpendicular to the jet for all epochs. 
VLBA observations of 3C\,345 at 8.4\,GHz made in 1997.07 (Taylor \cite{tay98})
yield an image that is very similar to our image at 1996.81, and both images 
imply a magnetic field aligned with the jet direction at 3--10\,mas distances
from the core.  Cawthorne \et\ (\cite{caw93}) suggested that 
the longitudinal component of the magnetic 
field can increase with the distance from the core as a result of
shear from the dense emission line gas near the nucleus.  At larger
distances from the core 
the shocks may become too weak to dominate the emission, resulting in
the observed electric field perpendicular to the jet.  This is possibly
the physical mechanism that can explain the phenomena observed in our
maps.


We also investigated the possible existence of a differential rotation
measure (RM) at the edges of the outer jet, comparing cuts across
the jet at 5\,GHz and 8.4\,GHz.  Such a RM would be caused by a
toroidal magnetic field (shear-like) in the jet (see Gabuzda \cite{gab99}). 
Owing to the uncertainties in $\chi$, the only reliable result is that 
the values of the EVPA are $\sim40^\circ$ bigger in the southern than 
in the northern edge, both at 5 and at 8.4\,GHz.  No conclusions
about the presence of this RM can be extracted.

\section{Summary\label{sec:discussion-summary}}

%
\begin{figure*}[tbhp]
\vspace{238pt}
\includegraphics{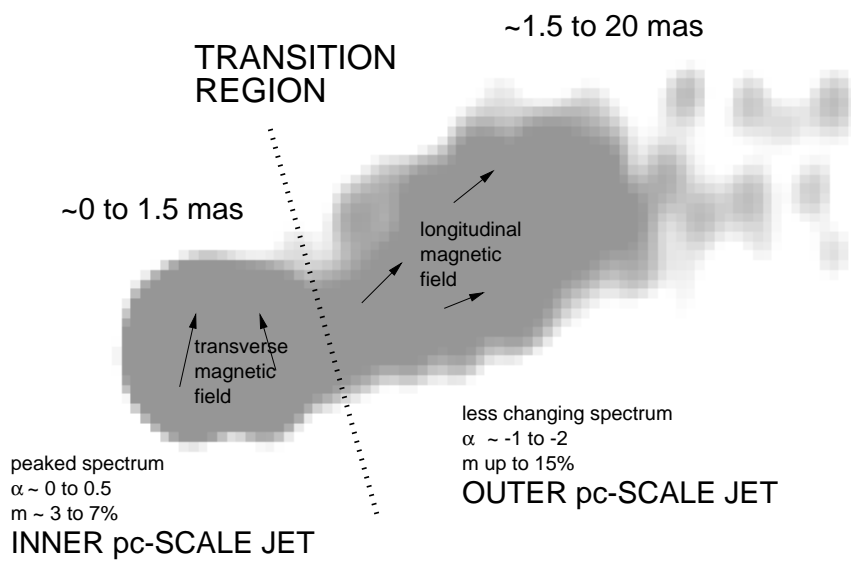}
\vspace{-25mm}
\hfill      \parbox[b]{5cm}{\caption[]{
Summary of physical properties observed in the two regions of the 
parsec-scale jet of \object{3C\,345}.
The  values for $\alpha$ and $m$ refer to 
our four observing frequencies (5, 8.4, 15, and 22\,GHz), in general.
\label{fig:summ}}
}
\end{figure*}

We have observed the superluminal QSO \object{3C\,345} with
VLBI at three epochs within one year,
observing with the VLBA at four frequencies.
Our results quantify the
superluminal motions of components C8 and C7 with respect to 
the core component D, and they show a remarkably complex polarization structure
near the core, which provides evidence
for emerging components and changing projected jet direction 
within 3\,mas from the core.
A comparison of our total intensity data at the four frequencies 
with previous model fitting results will provide a better understanding 
of the kinematical properties of the jet.

Spatial blending may be partly responsible for the low core polarization
observed for quasars in centimeter VLBI polarimetry.
Roberts \et\ 
(\cite{rob87}) mention the possible interference between components with
different $\chi$ in the polarized intensity 
for the case of \object{OJ\,287} at 5\,GHz.
Similar features can be present for the case of 
\object{3C\,345}, especially at 5\,GHz, since different
components have different values of $\chi$ at higher frequencies.

The polarized intensity images show alignment
between the jet direction and the
electric vector position angle in the inner regions of the jet at
high frequencies, and orthogonality of the electric vector and 
the jet direction in the outer regions of the jet.  These results
are fully consistent with the previously reported 
observations of Brown \et\ (\cite{bro94}), Lepp\"anen \et\ (\cite{lep95}),
Taylor (\cite{tay98}),
covering different frequencies and jet regions in \object{3C\,345}.

The interpretation of Lobanov \& Zensus (\cite{lob96b})
about a transition in the jet can be confirmed with the scenario
provided by our observations (see Fig.\ \ref{fig:summ}).  
The core region (up to 1.5\,mas), which presents
magnetic field orientations predominantly transverse 
(Figs.\ \ref{fig:pol22} \& \ref{fig:pol15-8.4}), would be dominated
by strong shocks (as components C8 and C7).  This shock dominance
is also consistent with the peaked spectrum (Fig.\ \ref{fig:maps-spix1.2}).  
At larger distances (beyond 1.5\,mas), the trajectories of the components are
less curved than in the inner jet
and show faster speeds, the magnetic field is longitudinal
(Fig.\ \ref{fig:pol5}) and the spectrum is smoother 
(Fig.\ \ref{fig:maps-spix1.2}).

The jet in 3C\,345, thus, apparently manifests different behaviors of the 
magnetic field,
which can be explained in terms of different mechanisms: (i) magnetic
field compression in the shocked, inner jet, (ii) ordering of the magnetic field
in a shear layer and edge-brightened polarized emission in the outer
jet, (iii) decoupling of two fluids within the jet with different properties,
obliqueness of shocks interacting with the outer edge of the jet (C8), etc.
To disentangle the different physical phenomena that are present in the
jet, further observations are needed, with the goal of higher resolution,
and including polarimetry.  The VSOP facility and the use of mm-VLBI
will provide important clues to the innermost structures of the
radio source (Klare \et\ in preparation).
Continuing the multi-band monitoring of \object{3C\,345} with VLBI, enhanced
by regular polarimetric studies, and also observations at
higher frequencies
and with better resolution (like those provided by orbital VLBI results
with HALCA) should help to better constrain the models of this QSO.

\begin{acknowledgements}
We want to acknowledge Drs.\ A.\ Alberdi and J.L.\ G\'omez for kindly 
providing D-terms information of the VLBA for epochs close to 1995.8
and 1996.8 from their observations of \object{4C\,39.25} and \object{3C\,120}.  We
acknowledge also Drs.\ R.W.\ Porcas, T.P.\ Krichbaum, and I.\ Owsianik,
for useful
discussions. We also want to acknowledge Dr.\ I.I.K.\ Pauliny-Toth for
a careful reading of the manuscript.
This research has made use of data of the
Michigan Radio Astronomy Observatory which is supported 
by the National Science Foundation and by funds from the University 
of Michigan.
The National Radio Astronomy
Observatory is a facility of the National Science Foundation operated
under cooperative agreement by Associated Universities, Inc.
\end{acknowledgements}


\begin{thebibliography}{}

\bibitem[1996]{all96}
Aller H.D., Aller M.F., Hughes P.A., 1996, In: Miller R., Webb J.R.,
Noble J.C.\ (eds), Blazar Continuum Variability, ASP 110, San Francisco,
CA, US, p.\ 208


\bibitem[1986]{bar86}
Bartel N., Herring T.A., Ratner M.I., Shapiro I.I., Corey B.E.,
1986, Nature 319, 733

\bibitem[1986]{bir86}
Biretta J.A., Moore R.L., Cohen M.H., 1986, 
\apj\ 308, 93


\bibitem[1994]{bro94} Brown L.F., Roberts D.H.,  Wardle J.F.C., 1994, 
\apj\ 437, 108 

\bibitem[1993]{caw93}
Cawthorne T.V., Wardle J.F.C., Roberts D.H., Gabuzda D.C., 1993, 
\apj\ 416, 519


\bibitem[1984]{cot84} Cotton W.D., Geldzahler B.J., Marcaide J.M.,
Shapiro I.I., Rius A., 1984, \apj\ 286, 503



\bibitem[1999]{gab99}
Gabuzda D.C., 1999, 
New Astronomy Reviews 43, 691

\bibitem[1994]{gom94}
G\'omez J.L., Alberdi A., Marcaide J.M., Marscher A.P., Travis J.P.,
1994, \aap\ 292, 33




\bibitem[1992]{hum92}
Hummel C.A., Schalinski C.J., Krichbaum T.P., 
\et,
1992, \aap\ 257, 489



\bibitem[1989]{kol89}
Kollgaard R.I., Wardle J.F.C., Roberts D.H., 
1989, \aj\ 97, 155

\bibitem[1995]{lep95} Lepp\"anen K.J., Zensus J.A., Diamond P.J., 1995, 
\aj\ 110, 2479 

\bibitem[1998]{lis98}
Lister M.L., Marscher A.P., Gear W.R.,
1998, \apj\ 504, 702



\bibitem[1996]{lob96}
Lobanov A.P., 1996, ``Physics of the Parsec-Scale Structures in the 
Quasar 3C\,345", PhD Thesis, New Mexico Institute of 
Mining \& Technology, Socorro, NM, US

\bibitem[1998a]{lob98a} 
Lobanov A.P., 1998a, 
\aap\ 330, 79 

\bibitem[1998b]{lob98b}
Lobanov A.P., 1998b,
\aaps\ 132, 261

\bibitem[1996]{lob96b}
Lobanov A.P., Zensus J.A., 1996,
In: Hardee P.E., Bridle A.H., Zensus J.A.\ (eds),
Energy Transport in Radio Galaxies and
Quasars, ASP 100, San Francisco, CA, US, p.\ 109

\bibitem[1999]{lob99} 
Lobanov A.P., Zensus J.A., 1999,
\apj\, 521, 509

\bibitem[1996]{mar96}
Marziani P., Sulentic J.W., Dultzin-Hacyan D., Calvani M.,
Moles M., 1996,
\apjs\ 104, 37




\bibitem[1988]{pea88}
Pearson T.J., Readhead A.C.S.,
1988, \aj\ 328, 114

\bibitem[1996]{qia96} 
Qian S.J., Krichbaum T.P., Zensus J.A., Steffen W., Witzel A., 1996, 
\aap\ 308, 395 

\bibitem[1995]{ran95}
Rantakyr\"o F.T., B{\aa}{\aa}th L.B., Matveenko L., 1995, 
\aap\ 293, 44

\bibitem[1987]{rob87} Roberts D.H., Gabuzda D.C., Wardle J.F.C., 
1987,
\apj\ 323, 536


\bibitem[1985]{rom85}
Romney J.D., 1985, VLBA Memo No.\ 452, NRAO, NM, US

\bibitem[1994]{she94}
Shepherd M.C., Pearson T.J., Taylor G.B.,
1994, BAAS 26, 987

\bibitem[1995]{ste95} 
Steffen W., Zensus J.A., Krichbaum T.P., Witzel A., Qian S.J., 1995, 
\aap\ 302, 335 



\bibitem[1998]{tay98}
Taylor G.B., 1998,
\apj\ 506, 637

\bibitem[1998]{ter98}
Ter\"asranta H., Tornikoski M., Mujunen A., \et, 1998,
\aaps\ 132, 305

\bibitem[1999]{tue99}
T\"urler M., Courvoisier T.J.-L., Paltani S., 1999,
\aap\ 349, 45

\bibitem[1983]{unw83}
Unwin S.C., Cohen M.H., Pearson T.J., \et,
1983, 
\apj\ 271, 536




\bibitem[1999]{vil99}
Villata M., Raiteri C.M., 1999,
\aap\ 347, 30

\bibitem[1994]{war94} 
Wardle J.F.C., Cawthorne T.V., Roberts D.H., Brown L.F., 
1994, 
\apj\ 437, 122 

\bibitem[1995a]{zen95a} 
Zensus J.A., Cohen M.H., Unwin S.C., 
1995a, 
\apj\ 443, 35 

\bibitem[1995b]{zen95b}
Zensus J.A., Krichbaum T.P., Lobanov A.P., 
1995b, Proc.\ Natl.\
Acad.\ Sci.\ USA 92, 11348

\bibitem[1998]{zha98}
Zhang X., Xie G.Z., Bai J.M., 1998,
\aap\ 330, 469

\end{thebibliography}
\end{document}